\RequirePackage{fix-cm}
\documentclass[smallextended]{svjour3}       
\smartqed  
\usepackage{graphicx}
\usepackage{latexsym}
\journalname{Experimental Astronomy}
\begin{document}
%
%
%


\def\icarus{\rm{Icarus}}
\def\aj{\rm{AJ}}                   
\def\araa{\rm{ARA\&A}}             
\def\apj{\rm{ApJ}}                 
\def\apjl{\rm{ApJ}}                
\def\apjs{\rm{ApJS}}               
\def\ao{\rm{Appl.~Opt.}}           
\def\apss{\rm{Ap\&SS}}             
\def\aap{\rm{A\&A}}                
\def\aapr{\rm{A\&A~Rev.}}          
\def\aaps{\rm{A\&AS}}              
\def\azh{\rm{AZh}}                 
\def\baas{\rm{BAAS}}               
\def\jrasc{\rm{JRASC}}             
\def\memras{\rm{MmRAS}}            
\def\mnras{\rm{MNRAS}}             
\def\pra{\rm{Phys.~Rev.~A}}        
\def\prb{\rm{Phys.~Rev.~B}}        
\def\prc{\rm{Phys.~Rev.~C}}        
\def\prd{\rm{Phys.~Rev.~D}}        
\def\pre{\rm{Phys.~Rev.~E}}        
\def\prl{\rm{Phys.~Rev.~Lett.}}    
\def\pasp{\rm{PASP}}               
\def\pasj{\rm{PASJ}}               
\def\qjras{\rm{QJRAS}}             
\def\skytel{\rm{S\&T}}             
\def\solphys{\rm{Sol.~Phys.}}      
\def\sovast{\rm{Soviet~Ast.}}      
\def\ssr{\rm{Space~Sci.~Rev.}}     
\def\zap{\rm{ZAp}}                 
\def\nat{\rm{Nature}}              
\def\iaucirc{\rm{IAU~Circ.}}       
\def\aplett{\rm{Astrophys.~Lett.}} 
\def\apspr{\rm{Astrophys.~Space~Phys.~Res.}}
\def\bain{\rm{Bull.~Astron.~Inst.~Netherlands}} 
\def\fcp{\rm{Fund.~Cosmic~Phys.}}  
\def\gca{\rm{Geochim.~Cosmochim.~Acta}}   
\def\grl{\rm{Geophys.~Res.~Lett.}} 
\def\jcp{\rm{J.~Chem.~Phys.}}      
\def\jgr{\rm{J.~Geophys.~Res.}}    
\def\jqsrt{\rm{J.~Quant.~Spec.~Radiat.~Transf.}}
\def\memsai{\rm{Mem.~Soc.~Astron.~Italiana}}
\def\nphysa{\rm{Nucl.~Phys.~A}}   
\def\physrep{\rm{Phys.~Rep.}}   
\def\physscr{\rm{Phys.~Scr}}   
\def\planss{\rm{Planet.~Space~Sci.}}   
\def\procspie{\rm{Proc.~SPIE}}   

\let\astap=\aap
\let\apjlett=\apjl
\let\apjsupp=\apjs
\let\applopt=\ao

\title{The Planet Formation Imager 
}

\titlerunning{Planet Formation Imager}        

\author{John D. Monnier, Stefan Kraus, Michael J. Ireland, 
 Fabien Baron,
 Amelia Bayo,    
 Jean-Philippe Berger, 
 Michelle Creech-Eakman,
 Ruobing Dong,
 Gaspard Duchene, 
 Catherine Espaillat, 
 Chris Haniff, 
 Sebastian H\"onig,  
 Andrea Isella,
 Attila Juhasz,  
 Lucas Labadie, 
 Sylvestre Lacour, 
 Stephanie Leifer,
 Antoine Merand,
 Ernest Michael,
 Stefano Minardi,
 Christoph Mordasini,  
 David Mozurkewich,
 Johan Olofsson,   
 Claudia Paladini,   
 Romain Petrov,
J\"org-Uwe Pott, 
 Stephen Ridgway, 
 Stephen Rinehart,
 Keivan Stassun,   
 Jean Surdej, 
 Theo ten Brummelaar,
 Neal Turner,  
 Peter Tuthill, 
 Kerry Vahala, 
 Gerard van Belle, 
 Gautam Vasisht, 
 Ed Wishnow,
 John Young,
 Zhaohuan Zhu
}

\authorrunning{Monnier, Kraus, Ireland, {\em et al}}

\institute{John D. Monnier \at
              1085 S. University Ave \\
              311 West Hall \\
              Ann Arbor, MI 48109 USA\\
              Tel.: +1-734-763-5822\\
              \email{monnier@Umich.edu}  
           }

\date{Received: date / Accepted: date}

\maketitle

\begin{abstract}
The Planet Formation Imager (PFI, www.planetformationimager.org) is a next-generation infrared interferometer array with the primary goal of imaging the active phases of planet formation in nearby star forming regions.  PFI will be sensitive to warm dust emission using mid-infrared capabilities made possible by precise fringe tracking in the near-infrared.  An L/M band combiner will be especially sensitive to thermal emission from young exoplanets (and their disks) with a high spectral resolution mode to probe the kinematics of CO and H$_2$O gas.  In this paper, we give an overview of the main science goals of PFI, define a baseline PFI architecture that can achieve those goals, point at remaining technical challenges, and suggest activities today that will help make the Planet Formation Imager facility a reality. 
\end{abstract}

\section{Introduction}
\label{intro}

Planet formation is one of the most exciting areas of modern astronomy.  A complete theory of planet formation promises to bridge the fields of star formation with exoplanets, seeking to understand the observed distribution of exoplanets as function of orbital radius, stellar mass, metallicity, binarity, etc.  With a well-validated model, we could confidently simulate aspects of planet formation that are difficult to directly observe, such as the distribution of water and the likelihood that exoplanets have environments suitable for life.  

Using spectral energy distributions as a guide, astronomers \cite{smith1984,calvet2002,espaillat2011} have identified young stars with disks in varying stages of planet formation, showing inner dust holes, large gaps, and everything in between.  Interpreting these observations without spatially-resolved imaging has proven difficult and recent imaging advances using adaptive optics coronagraphy \cite{macintosh2008,beuzit2008} and mm-wave imaging \cite{alma2015} has revealed a rich set of phenomena.   The presence of spirals, multiple rings, and even more complex structures of dust and gas serve surely to reveal key details of the planet formation process.  
With current techniques limited to typical spatial resolutions of $\sim5$-$10$ astronomical unit (au) in nearby star-forming regions, we must look to infrared interferometry to reach the sub-au spatial scales (MILLI-arcsecond angular scales) sufficient to resolve disk gaps cleared by single planets, to detect accretion streams, and to follow the dust and gas all the way down to scales of individual exoplanet Hill Spheres ($\sim$0.03 au), where disk material is accreted onto young planets themselves.    

An effort was started in 2013 at a workshop in Haute-Provence to explore the potential for a new facility that could image all the fundamental stages of planet formation {\em in situ} in nearby star forming regions.  The Planet Formation Imager (PFI, www.planetformationimager.org) project was first proposed at the 2014 SPIE meeting on Astronomical Telescopes and Instrumentation (Montreal) in a series of presentations \cite{pfimonnier2014,pfikraus2014,pfiireland2014}.

Here we summarize the progress the PFI teams have made in defining top-level science requirements and a baseline facility architecture for achieving these goals.  We also highlight the importance of technical progress needed to make PFI more affordable and point out how existing interferometer facilities can play a critical role in development key PFI technology.
\begin{figure*}
  \includegraphics[width=1.0\textwidth]{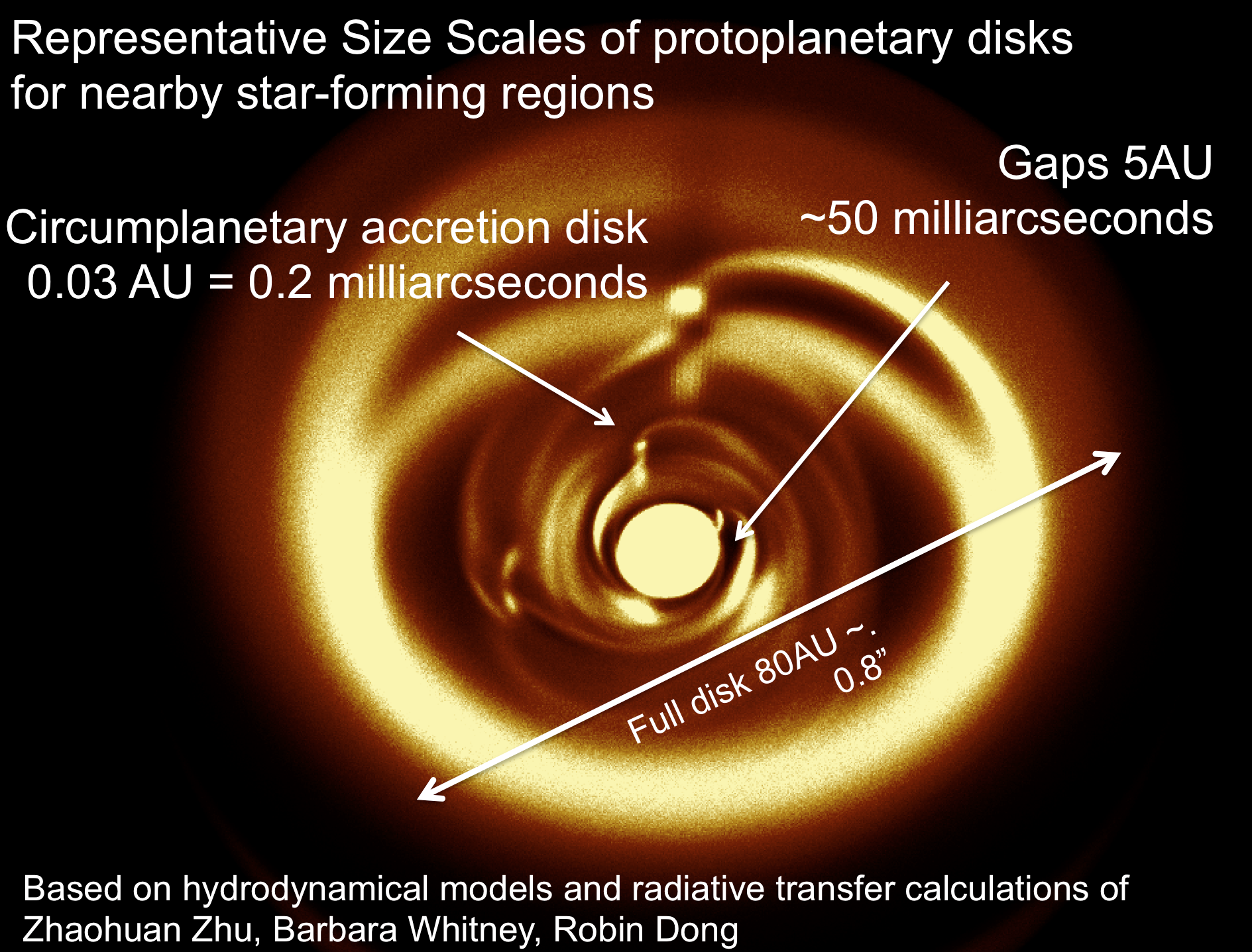}
\caption{Radiative transfer model for an example planet-forming disk \cite{pfimonnier2014,dong2015} with the relevant size scales marked. The primary science driver of the Planet Formation Imager (PFI) is to image scales as large as the whole circumstellar accretion disk down to the circumplanetary accretion disks of individual giant planets.  
Simulation assumptions: M$_\ast=1.0\,M_\odot$, T$_\ast=4500\,K$, R$_\ast=2.0\,R_\odot$, M$_{\rm disk}=0.04\,M_\odot$, 4x2\,M$_{\rm Jup}$ planets.}
\label{fig:1}       
\end{figure*}

\section{Science Goals of PFI}

We introduce the science goals of PFI by looking at a radiative transfer image of a planet-forming disk in the thermal infrared (see Figure~\ref{fig:1}).  The protoplanetary disk is approximately 100\,au across, with gaps and structures on the scale of $\sim$5\,au.  We expect a circumplanetary disk to form on scales of 0.03\,au, matching the Hill Sphere for each accreting protoplanet (for a Jupiter-mass planet on a 5\,au radius orbit).  The mid-infrared wavelength range efficiently traces emission from small grains from 0.1-10\,au in the disk, complementing mm-wave/radio observations of the large grains.  In the mid-infrared, probing scales of 0.1\,au at the distance of the most nearby star forming regions would require a telescope with diameter much larger than 100\,m and so we focus on infrared interferometry with kilometric baselines as a facility architecture.

In order to define a set of top-level science requirements, we must consider the relevant fluxes of young stars, their circumstellar planet-forming disks, their orbiting young exoplanets, and their warm circumplanetary accretion disks.  In turn, these fluxes will drive our technical architectures, such as the size and number of telescopes needed in the array. We used models from Baraffe et al. \cite{baraffe1998} to estimate fluxes from a typical T~Tauri star.  For the young exoplanets themselves, we considered hot and cold-start models from Spiegel et al. \cite{spiegel2012}.  While estimates for the luminosity and temperature structure of circumplanetary accretion disks are very uncertain, Zhu \cite{zhu2015} provided the information we used for the circumplanetary accretion disks here.  Lastly, the dust structures, densities, and temperatures were provided from a 4-planet disk model calculated by Dong et al. \cite{dong2015} (Figure~\ref{fig:1}).  We are working with hydrodynamicists to produce more simulations that cover a range of disk ages and planet-formation scenarios to better understand the scope of the achievable science cases.

Table~\ref{relevantfluxes} summarizes these critical parameters that are key for driving the top-level science requirements.  Note that an adaptive optics system is essential to use the full aperture of the telescope for interferometry, imposing flux limits in the R/I band for wavefront sensing.  
  
\begin{table}[ht]
\caption{Typical absolute magnitudes for the emission components in protoplanetary disks \cite{pfimonnier2016,baraffe1998,spiegel2012,zhu2015,dong2015}, for the wavelength bands relevant for PFI, including adaptive optics system (Y band), fringe tracking system (H band), young exoplanets and dusty structures in the disk (L and N band).
To convert these absolute magnitudes to apparent magnitudes for an object located in Taurus at 140\,pc, simply add 5.7 magnitudes to the numbers below. }
\label{relevantfluxes}
\begin{center}       
\begin{tabular}{|l|c|c|c|c|}
\hline
Component & $M_Y$ & $M_H$ & $M_L$ &$M_N$ \\
          & (AO & (fringe & (dust & (dust \\
          & system)  & tracking) & \& planets) &  \& planets)\\
\hline
Example T Tauri Star & & & &  \\
\qquad 1 M$_{\odot}$, 2.1\,R$_{\odot}$, 3865\,K& 4.9 & 2.54 & $\sim$2.5  & $\sim$2.4 \\
\qquad 3\,Myr, [Fe/H]=0.0 &&&&\\
\hline
Protoplanet  & & & &  \\
\qquad ``hot start", 2M$_J$, 1Myr & & 12.9 & 11.0 & 9.1  \\
\qquad ``cold start", 2M$_J$, 1Myr& & 18.2 & 14.7 & 11.2   \\
\hline
Circumplanetary Disk & & & & \\
\qquad ($R_{in}=1.5~M_J$) & & & & \\
\qquad $ M\dot{M}=10^{-6}M^2_J\,{\rm yr}^{-1}$ & & 16.4 & 9.8 & 6.5  \\
\hline
4-planet gapped disk & & & &\\
\qquad Star only (2\,R$_{\odot}$, 4500\,K) & 4.1 & 2.1 & 2.1  & 2.1  \\
\qquad Star + Disk (30$^\circ$ inclination)                & 4.1 & 2.1 &  1.6 & -1.1  \\
\hline
\end{tabular}
\end{center}
\end{table}

Having reviewed the typical characteristics of key science targets, we have summarized our top-level science requirements in Table~\ref{tlsr}.  There are hundreds of young stellar objects with disks that satisfy these requirements within 200~pc and thousands if we move out to the distance of 1~kpc.  In the next section, we propose a specific facility architecture that can achieve most of the top level science requirements at the cost of a typical major astronomical facility.

We note that our science working group is also very interested in science cases beyond planet formation. Imaging dust tori around Active Galactic Nuclei, stellar orbits around the Galactic Center, AGB stars mass-loss, diameters of young stars themselves, magnetic spots on main sequence stars, and more are all possible with PFI -- but the baseline design is focused on the planet formation case to define the required facility architecture.

\begin{table}
\begin{center}
\caption{Top-level Science Requirements (Minimum Goals)}
\begin{tabular}{|l|c|c|}
\hline
 Parameter & Dust Imaging & Young Exoplanets\\
\hline
Wavelengths & 5-13\,$\mu$m & 3-5\,$\mu$m \\
Typical Source Distance & 140\,pc & 50-500\,pc \\
Spatial Resolution & 2\,mas $\equiv$ 0.3~AU & 0.7\,mas $\equiv$ 0.1~AU (for 140pc)\\
Point-Source Sensitivity & $m_N\sim12.5$ (5-$\sigma$)  & $m_L\sim18.5$ (5-$\sigma$) \\
\qquad (t$=10^4$s) & & \\
Goal Surface Brightness (K) & 150\,K  &  $--$\\
\qquad (t$=10^4$s) & & \\
Spectral Resolving Power&&\\
\qquad Continuum      & R$>100$  & R$>100$ \\
\qquad Spectral Lines & R$>10^5$ & R$>10^5$ \\
Field-of-view &  $>0.15"$  & $>0.15"$ \\
Minimum Fringe Tracking Limit & $m_H<9$ (star only) & $m_H<9$ (star only)\\
Fringe tracking star & $\phi<0.15\,$mas & $\phi<0.15\,$mas \\
\hline
\end{tabular}
\label{tlsr}
\end{center}
\end{table}

\section{Technical Description of the PFI Array}

After the 2014 SPIE meeting where the PFI project was introduced, a Science Working Group (SWG; headed by Stefan Kraus) and a Technical Working Group (TWG; headed first by David Buscher, and now Michael Ireland) were formed involving around one hundred astronomers around the world.  Based on the early top-level science requirements first outlined in 2014, the 2016 SPIE meeting in Edinburgh saw even more contributions which explored technical solutions to achieve these science goals \cite{pfimonnier2016,pfikraus2016,pfiireland2016,pfiminardi2016,pfimozurkewich2016real,pfibesser2016,pfipetrov2016}.  For instance, a mid-infrared wavelength range was chosen over mm-wave or near-infrared to access the most diverse aspects of planet formation in the "warm dust" zone.  We also seriously explored a new heterodyne interferometer concept using mid-IR laser combs \cite{pfiireland2014,pfibesser2016}, although our current baseline array now favors direct detection method to allow L/M band wavelengths which are key to detect young protoplanets themselves.  The infrared surface brightness sensitivity is mostly determined by the size of the individual apertures and not the number of telescopes -- this pushed the design towards large-area unit telescopes which drives the cost.  A simple cost model was introduced by Ireland et al.\ \cite{pfiireland2016} which informed the baseline architecture described now where fewer large telescopes were preferred over many more small apertures (at fixed cost).  

\subsection{Baseline Architecture}

Table~\ref{pfispecs} contains a summary of a baseline PFI architecture sufficient to achieve the top-level science requirements.  PFI here consists of twelve 3-m class telescopes arranged in either a "Y-array" or "ring array" with maximum baselines of 1.2\,km.   With this geometry, fringe tracking can be done using the shorter spanning baselines while the longest baselines provide an angular resolution of 0.6 / 1.7 milliarcsecond resolution at L band (3.5$\mu$m) / N band (10$\mu$m), which corresponds to spatial resolution of 0.08\,au / 0.25\,au at 140\,pc.  The L/M band angular resolution and a high spectral resolution mode should allow spectro-astrometric detection of CO/H$_2$O gas kinematics in circumplanetary disks when bright enough.  Reliable calculations of the molecular gas emission are not yet available and are being pursued by our Science Working Group to evaluate this science goal more soberly.

\begin{table}
\begin{center}
\caption{Technical Description of Baseline PFI Architecture}
\begin{tabular}{|l|c|}
\hline
Parameter & Range \\
\hline
Number of Telescopes & 12 \\
Telescope Diameter & 3\,m \\
Maximum Baseline & 1.2\,km \\
Goal Science Wavelengths & 3--13\,$\mu$m \\
Fringe-tracking wavelengths & 1.5--2.4$\mu$m \\
Fringe tracking limits & $m_H<$13 for point source \\
Field-of-view & 0.25" (L band), 0.7" (N band) \\
Construction cost & \$250M\\
\hline
\end{tabular}
\label{pfispecs}
\end{center}
\end{table}

Our initial simulations of the $10\,\mu$m imaging performance of PFI \cite{pfimonnier2016} show that even a 20-element 2.5m telescope array lacked sufficient sensitivity to see the lower surface brightness and smallest disk features with the conservative assumption of astronomical silicates as the main dust type.  Because of this, the baseline architecture presented here chose larger telescopes (3\,m) but had to reduce the number of telescopes to 12 with maximum baseline 1.2\,km (rather than the 5-10\,km baselines originally).  In the next section we outline the PFI technical roadmap, which has the top priority to investigate new telescope technologies to allow $5+$ meter telescopes at low enough cost to populate a 20$+$ element array.

The PFI SWG and TWG will carry out more sophisticated modelling of a more diverse set of planet-forming disk simulations to explore what disk structures can be reliably imaged for various technical assumptions.  In the context of realistic simulations, our team can explore whether extensions of PFI toward Q-band (18$\mu$m) might be warranted.


\subsection{Potential Sites}
The PFI SWG determined that a mid-latitude site is near-essential for PFI due to the limited number of star-forming regions observable from high-latitude sites, which removes the High Antarctic Plateau from consideration. The PFI Project has identified the Flagstaff (Arizona, USA) Navy Precision Optical Interferometer (NPOI) site and the ALMA site (Chajnantor Plateau, Chile) as locations with sufficient accessible area and existing infrastructure to merit further consideration.

\section{On the Path to PFI}
\subsection{Key PFI Technologies}

\begin{figure*}
  \includegraphics[width=.55\textwidth]{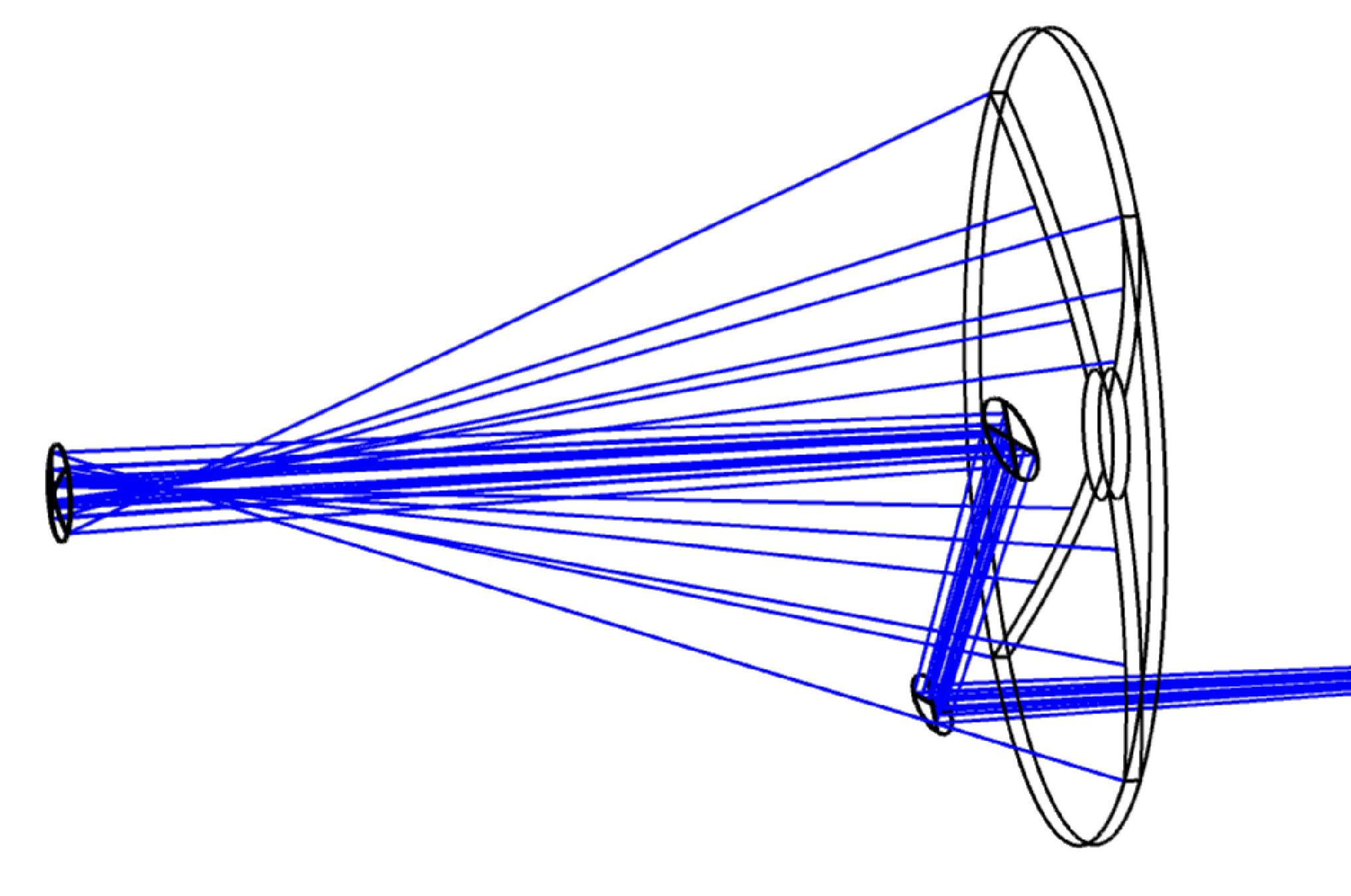}
    \includegraphics[width=.45\textwidth]{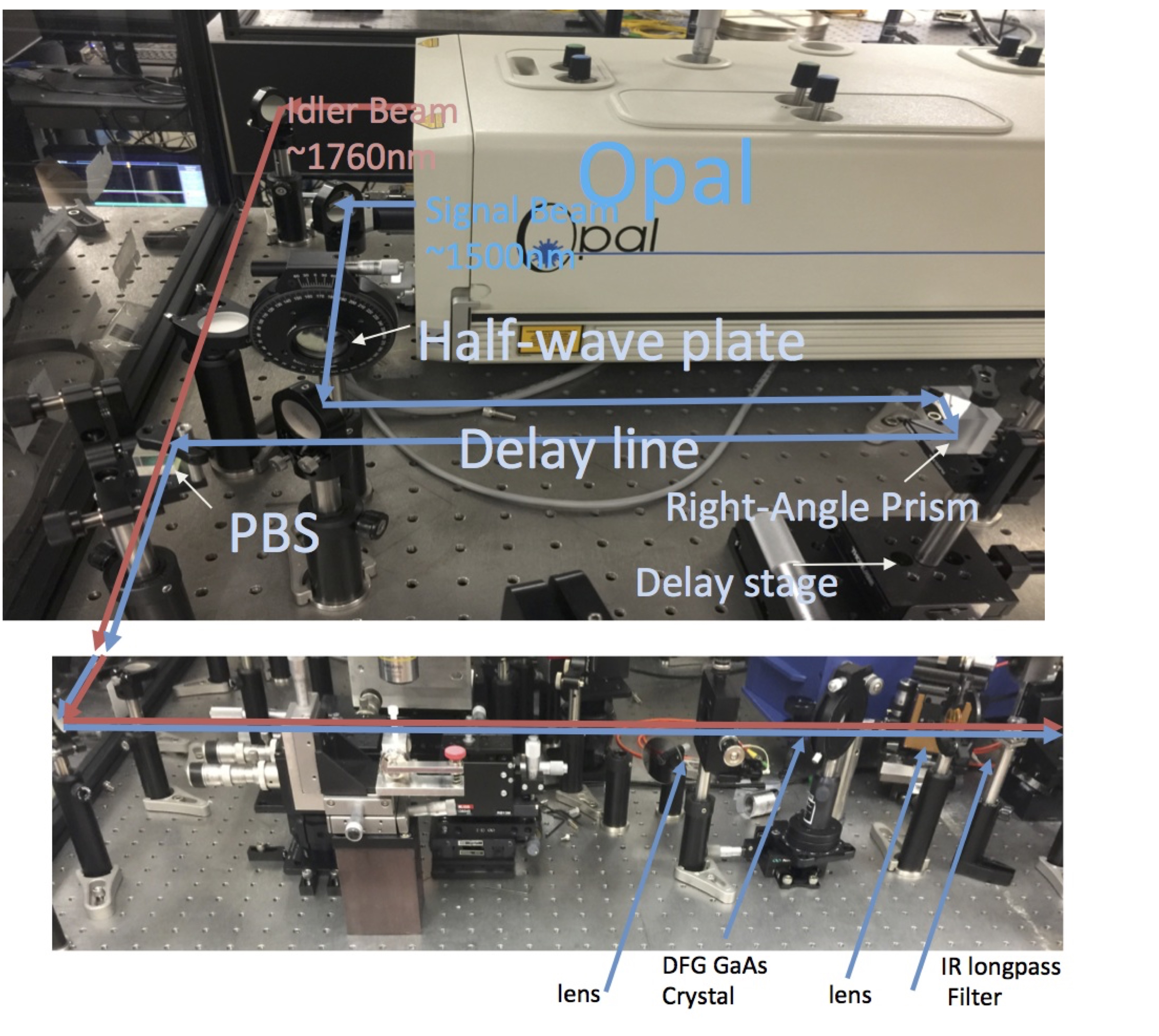}
\caption{(left) A spherical primary can be corrected for diffraction-limited imaging using an aspherical secondary, although the corrected field-of-view is extremely small.  By using a Gregorian design, the pupil is apodized to be more gaussian, ideal for both propagation through long delay lines and injection into a single-mode fiber.  See Monnier et al. (2014) and Ireland et al. (2016) for more information \cite{pfimonnier2014,pfiireland2016}.
(right) Photograph shows the laboratory set-up to explore the generation of N-band laser frequency combs using direct difference frequency generation (DFG) of 
near infrared femtosecond pulses output by an optical parametric oscillator. Note that no N-band light has been generated yet and this is a work in progress as a
collaboration between JPL and Caltech Applied Physics. Other approaches include direct LFC generation \cite{hugi2012}, while recent results on 
intrapulse DFG for this frequency range are found in Timmers et al (2017) \cite{timmers2017}.}
\label{fig:tel}       
\end{figure*}

The PFI TWG has identified key technologies needed to lower technical and financial risk before a facility is constructed.  For science performance, the PFI benefits directly from larger telescopes but telescope costs scale steeply with aperture, making telescope construction the main cost driver for PFI.  Multiple avenues are being explored to ``break'' the existing cost scaling curve for large telescopes, including spherical primaries, carbon-fiber reinforced polymers for mirror replication and/or for lightweight supports, and more. Natural allies for this technical development include governments interested in imaging geostationary satellites and telecommunication companies interested in narrow-field, diffraction-limited applications such as laser communication (to/from space or ground stations).  Work is ongoing in Chile (Valparaiso), USA (Flagstaff, Michigan), and Australia (Canberra) to seek funding and new partnerships.  Figure~\ref{fig:tel} shows a ZEMAX design for a spherical primary with a highly-aspheric Gregorian secondary resulting in a gaussian-like apodization of the pupil with diffraction-limited performance over a small field-of-view \cite{pfimonnier2014,pfiireland2016}.  The field-of-view limitations of some designs would limit off-axis AO tracking for visibly-faint targets and this trade-off requires more study.

Beam combination technologies need further development as well, especially at L/M band where integrated optics technologies could make beam combination simultaneously less expensive, vastly simpler, and with better calibration.  Exciting work is going on in the UK, Germany, Belgium, France, and Australia.  The HI-5 experiment to develop L-band nulling on the VLTI would be an exciting pathfinder for PFI.

While the Keck Interferometer Nuller \cite{colavita2010} and the VLTI/MIDI+FSU experiment\cite{muller2014} showed that precise fringe tracking in the near-infrared can track piston variations in the mid-infrared, the upcoming "GRA4MAT" mode of the VLTI will phase the 4-telescope L/M/N-band MATISSE instrument using the K-band GRAVITY instrument fringe tracker.  This mode will allow precise testing of algorithms and validate atmospheric modelling to an unprecedented level, and act as a kind of PFI testbed.

A high-sensitivity fringe tracker should be built and tested to validate the PFI instrument modelling and support our $H\sim$13 magnitude limit (for 3-m AO-corrected telescopes) estimate for fringe tracking performance.  Perhaps such a system could be paired with a new J-band combiner to carry-out a large  stellar multiplicity survey based on upcoming GAIA survey results, either at MROI, CHARA, NPOI, or VLTI.

Highly-multiplexed heterodyne interferometry based on new laser combs could be competitive with the "direct detection" scheme presented here.  Heterodyne has an advantage for a large number of telescopes ($N>40$) and if one can avoid vacuum beam delay lines (for instance, by using fiber optics for fringe tracking wavelengths); however, the need for fewer large telescopes to reduce IR background and the scientific focus on L,M band (requiring vacuum beam transport) both act to negate the exciting advantages of the heterodyne approach.  That said, the PFI TWG supports continued work, such as that going on in Chile and at JPL, and we hope to hear more at the 2018 SPIE.  Figure~\ref{fig:tel} shows a lab prototype N-band laser comb under development at JPL with Caltech Applied  Physics.

PFI should also seek to operate in a stream-lined, efficient, and low-cost manner. The Palomar Testbed Interferometer (PTI) \cite{colavita1999} was a famously efficient system which collected data on thousands of objects over many years with a low operations budget.  PFI should be considering operations modelling as planning moves towards a Phase A study.

We have summarized these roadmap recommendations in Table~\ref{tab:roadmap}.

\subsection{A Pathfinder Facility?}

A new interferometric array of limited scope could be a powerful vehicle to debut the core technologies needed to build the full PFI. For instance, a 3-telescope array using new 3-5\,m class telescopes with kilometric baselines could test the new "cheap telescopes" in practice, test kilometer-length delay line technology, validate high-throughput beam train design, commission a high-sensitivity fringe tracker, and adopt a low-cost operations model.  Such an instrument could survey 10000+ binary systems with astrometric and RV orbits from GAIA and other large scale surveys.  By resolving these binaries with an interferometer at even a single epoch, masses can be determined for all components, unlocking powerful avenues to probe stellar structure and evolution with powerful application to Galactic archeology.  While a 3-telescope array would not be able to image details of complex sources such as YSO disks or AGB star photospheres, this high sensitivity and high angular resolution facility would open other new scientific areas such as testing pre-main sequence stellar evolution through diameters of young stars and could finally clearly resolve near-IR AGN for the first time.  Such a facility -- Stellar Multiplicity Interferometer Large Experimental Survey (SMILES) -- could be a powerful achievable goal during the next decade. Further, if built at a location compatible with the full PFI, some of the SMILES infrastructure could potentially become part of PFI.

\begin{table}[ht]
\caption{PFI Technology Roadmap} 
\label{tab:roadmap}
\begin{center}       
\begin{tabular}{|l|l|}
\hline
Critical Technology &  Considerations \\
\hline
Inexpensive telescopes & Possible key technologies: \\ 
& Replicated parabolic lightweight mirrors,\\
& inexpensive primaries with high low-order errors,\\
& lightweight structures with exquisite AO controls, \\
& {\bf Partner with industry, engineers}\\
& {\bf New telescopes tested on existing arrays}\\
\hline
L/M band IO combiners 
& Needed for high precision calibration, \\
& Explore Chalcogenide integrated optics, \\
& {\bf Pilot project at VLTI such as HI-5 } \\
\hline
Wavelength-bootstrapped  
& L band imaging require $10^6$:$1$ dynamic range imaging, \\
fringe tracking 
& ultra-accurate fringe tracking in L based on H-band, \\
 & {\bf VLTI/GRA4MAT mode}\\
& while maintaining very high sensitivity, \\
& {\bf New "high sensitivity" fringe tracker at VLTI}\\
\hline
Mid-IR laser comb heterodyne & Possible "add-on" to L/M band \\
 & {\bf Develop combs, detectors, digital processing} \\
\hline
Low-cost operations model  & {\bf New array of limited scope, } \\
& {\bf e.g., a Stellar Multiplicity 3-telescope Experiment} \\
\hline
\end{tabular}
\end{center}
\end{table}

\section{Conclusion}

The Planet Formation Imager Project has spent the last 4 years defining realistic science goals and a practical facility to achieve them.  Actually imaging the major stages of planet formation -- {\em as they are happening live} -- is a realizable dream using today's infrared interferometric technologies.  New technologies could lower the cost of PFI and make the capabilities even more powerful over the next decade.  In addition to purely technological development, we have also identified pathfinder instruments and a possible new limited-scope facility that can yield exciting short/medium-term science while laying the groundwork for the ambitious PFI facility in the next decade.

\begin{acknowledgements}
The PFI project team would like to acknowledge Professor Charles Townes for his profound impact on the field of stellar interferometry – Dr. Townes passed away January 27, 2015 and will be missed by many in our field. 
\end{acknowledgements}

\bibliographystyle{spmpsci}      
\bibliography{pfi1}   

%
%

\end{document}